\documentstyle[12pt]{article}
\def\baselinestretch{1.5}
\begin{document}
\thispagestyle{empty}
\null\vskip -1cm
\centerline{
\vbox{
\hbox{March 1997}\vskip -9pt
\hbox{hep-ph/9703435}\vskip -9pt
     }
\hfill
\vbox{
\hbox{NHCU-HEP-97-5}
\hbox{UICHEP-TH/97-3}
     }     } \vskip 1cm

\centerline
{\large \bf
Electron Electric Dipole Moment from }
\centerline
{\large \bf CP Violation in
the Charged Higgs Sector
}
\vspace{1cm}
\centerline           {
David Bowser-Chao$^{(1)}$,
Darwin Chang$^{(2,3)}$,
and
Wai-Yee Keung$^{(1)}$ }
\begin{center}
\it
$^{(1)}$Physics Department, University of Illinois at Chicago,
IL 60607-7059, USA\\
$^{(2)}$Physics Department,
National Tsing-Hua University, Hsinchu 30043, Taiwan, R.O.C.\\
$^{(3)}$Institute of Physics, Academia Sinica, Taipei, R.O.C.\\
\vspace{.5cm}
\end{center}

\begin{abstract}
The leading contributions to the electron (or muon) electric dipole
moment due to CP violation in the charged Higgs sector are at the
two--loop level.
A careful model-independent analysis of the heavy fermion
contribution is provided.
We also consider some specific scenarios to
demonstrate how charged Higgs sector CP violation can naturally give
rise to large electric dipole moments.
Numerical results show that the
electron electric dipole moment in such models can lie at the
experimentally accessible level.

\end{abstract}

\vspace{1in}
\centerline{PACS numbers:  13.40.Fn, 11.30.Er, 12.15.Cc, 14.80.Dq \hfill}
\newpage

\section*{Introduction}


\quad
\vskip -1cm

Experiment has established that neither parity (P) nor charge
conjugation (C) are unbroken symmetries. Kaon
physics show that CP also fails to be an exact symmetry. The
CPT theorem then implies that time-reversal (T) is broken as
well, leading to expectation of a T-odd electric dipole moment (EDM)
for one or more of the elementary particles.
The Standard Model (SM) of electroweak interactions explains CP
violation in the $K-{\bar K}$ system as the result of a single complex
phase in the Cabibbo-Kobayashi-Maskawa (CKM) matrix. It also predicts an
electron EDM $d_e$ of about $8 \times 10^{-41} {\rm e} \cdot {\rm cm}$
\cite{SM}
and a muon EDM of about $2\times 10^{-38} {\rm e}\cdot
{\rm cm}$, while the neutron EDM (calculated from the up and down quark
EDMs) is estimated  \cite{nedm} to be less than $10^{-31} {\rm e}\cdot
{\rm cm}$,
The experimental limits (given at  95\% C.L.) are several orders of
magnitude above these predictions, with the limit on the electron EDM
$|d_e| < 6.2 \times 10^{-27} {\rm e}\cdot {\rm cm}$ \cite{de}. The limit
on $d_\mu$ is even further removed from the SM prediction, with $|d_\mu|
< 1.1 \times 10^{-18} {\rm e}\cdot {\rm cm}$ \cite{dmu}, 
although there is a proposal to measure
the muon EDM down to $10^{-24}$ e$\cdot$cm \cite{E821}.
The neutron EDM limit is $|d_n| < 11 \times 10^{-26} {\rm e}\cdot {\rm
cm}$ \cite{dn}. Clearly, measurement of a non-zero electron, muon, or
neutron EDM close to current or proposed limits would point to physics
beyond the Standard Model.

New sources of CP violation can come from complex couplings or vacuum
expectation values (VEV) associated with the Higgs boson sector. A
significant EDM for elementary fermions can be
generated if CP violation is mediated by neutral Higgs-boson exchange
\cite{Steve89,Steve90,Barr&Zee,BZfollow,Barger}. Dominant contributions
come from one-loop or two-loop diagrams. The one-loop terms are
proportional to $(m/v)^3$ (with one factor of $m/v$ due to an internal
mass insertion), while the two-loop terms are
proportional\cite{Barr&Zee} to $m/v$, with $m$ being the fermion mass
and $v = 246$ GeV the vacuum expectation value of the SM Higgs field.
The one-loop contributions are thus strongly suppressed relative to the
two-loop terms, by a factor of $(m/v)^2$.
Exhaustive studies\cite{BZfollow,Kao&Xu} have been carried out at the
two loop level on the electron EDM generated by CP violation in the neutral
Higgs-boson sector.
The contributions containing a heavy fermion or gauge boson loop
were considered\cite{BZfollow}
and, within an $SU(2) \times U(1)$ theory, are more or less
model-independent.  Additional, relatively more model-dependent,
contributions involving a physical charged
Higgs-boson loop (with a CP  violating charged-Higgs-neutral-Higgs
coupling) have been considered in Ref.\cite{Kao&Xu}. On the other hand,
the corresponding contribution to $d_e$ due {\em solely}
to CP violation in charged Higgs sector has not been studied in
the literature, even though this category of CP violating mechanism has been
emphasized for other phenomenological effects such as
top quark decay\cite{ch-top}, the neutron electric dipole
moment\cite{CKLY,ch-nedm}, and $\Gamma(b\to s \gamma)$\cite{ch-bsg}.

With charged Higgs sector CP violation involving fermions, the one-loop
contribution is
suppressed as in the neutral Higgs case  but suffers an additional
factor of $m_\nu/m_e$, where $m_\nu$ is the mass of the electron
neutrino.
If no right-handed neutrino exists, or the neutrino is
massless, the two-loop diagrams are unequivocally the leading
contribution. It is also important to note that the recent measurement
of the decay rate of $b \to s\gamma$ by the CLEO collaboration
\cite{CLEO} stringently constrains\cite{bsg} the mass of the $H^\pm$
only in the 2HDM II. The constraint is easily evaded in 2HDM
III\cite{ch-bsg,Wolf,Bowser} and other extensions.

In this letter, we make a model-independent study of the two--loop
contribution to $d_e$ and $d_\mu$
from charged Higgs exchange between fermions.
Useful formulas are given.  We
discuss specific models to see how  this type of CP
nonconservation can arise.
Numerical results show the electron electric
dipole moment can naturally lie within reach of  experiment.
We also
present results for the corresponding contributions to
$d_n$.

\section*{General Formalism}

\quad \vskip-1cm

We first comment on the model-independence of our analysis,
which considers only those graphs that involve CP violation from
charged Higgs exchange between
fermions. There are other possibilities,
such as quartic charged Higgs vertices involving three or four distinct
charged Higgs bosons with complex coupling, but these are strongly
dependent on model details and less amenable to
model-independent parameterization.
Secondly, one may have several charged Higgs bosons,
but barring significant degeneracy,
cancellation among contributions from the various
charged Higgs should be mild. For simplicity we only consider
contributions from the lightest charged Higgs.
We shall ignore model dependent neutral
Higgs sector CP violation contributions and
CP violation involving both neutral and charged Higgs boson,
as our concern here is with exclusively
charged Higgs sector CP violation effects.
We shall also omit terms
where the neutral Higgs bosons participate only incidentally, serving as
a leg of an internal loop but not contributing a CP phase.
We expect
that these (perhaps  non-negligible) contributions
should not strongly cancel with the part studied here; thus our
analysis should furnish a reasonable lower limit to $d_e$ for general
theories with CP violating charged Higgs exchange between fermions.

We parametrize the charged Higgs sector CP violation as follows:
$$
{\cal L}={g\over\sqrt{2}}                         \left(
         {m_t c_t\over M_W}
          \bar t_R b_L H^+
        + {m_e c_e\over M_W}
          \bar \nu_L e_R H^+
        + \bar t_L\gamma^\alpha b_LW^+_\alpha
        + \bar \nu_L\gamma^\alpha e_LW^+_\alpha   \right)
$$
\begin{equation}
        -m_e \bar{e_L}e_R -m_t \bar{t_L}t_R + {\rm H.c.}
\label{eq:lag}
\end{equation}
Note that in our model-independent analysis, one needs
not specify the origin of the CP violation (e.g., explicit or
spontaneous CP violation).

We only illustrate the most important
contribution from the top-bottom generation; our study can easily be
generalized to the three generations case. The bottom quark mass $m_b$
is also set to zero. The
complex mixing parameters $c_t$ and $c_e$ signal deviations from the
2HDM II. If $c_t c_e^*$ has a non-zero imaginary part, the phase is
intrinsic to the lagrangian and cannot be rotated away by redefinition
of any or all of the fields in Eq.(\ref{eq:lag}).

\begin{sloppypar} The two-loop charged Higgs contribution involves
Feynman diagrams such as the one shown in Fig.~1. We first  present a
simple expression for the one-loop sub-diagram with fermion in the loop,
that is, the trun\-cated three-point Green's func\-tion $\Gamma^{\mu\nu}
= \langle 0 |[H^{-}(p) A^\mu(k) {W^{+}}^\nu(-q)]_+|0\rangle$. We note
that $\Gamma^{\mu\nu}$ is the off-shell extension of the amplitude
(and Feynman diagrams) for
$H^-\to W^- \gamma$ given in Ref.\cite{Raychaudhuri}, and that both
$\Gamma^{\mu\nu}$ and its charge conjugate contribute to $d_e$.
We consider only
the (gauge-invariant) set of terms
in $\Gamma^{\mu\nu}$ involving CP violation from
charged Higgs exchange between fermions.
\end{sloppypar}

We have verified that our results are gauge-independent,
but the calculation simplifies greatly
in the non-linear $R_\xi$ gauge\cite{Gavela}. Since the EDM is defined
in the soft photon limit, only the leading term in $k$ is kept. We also
work to lowest
order in $(m_e/M_W)$, to which approximation the separation of
the calculation into $\Gamma^{\mu\nu}$ and its insertion in the full
two-loop graph is gauge-invariant. We thus obtain:

\begin{equation}
\Gamma^{\mu\nu}
={3eg^2\over 16\pi^2} { c_t \over M_W}
 [( g^{\mu\nu}q\cdot k -q^\mu k^\nu)S
 +iP\varepsilon^{\mu\nu\alpha\beta}p_\alpha q_\beta] \ ,
\end{equation}

\begin{equation}
S=  \int_0^1 {q_t(1-y)^2+q_b y(1-y) \over 1-yq^2/m_t^2}dy
\ , \quad
P=  \int_0^1 {q_t(1-y)+q_b y \over 1-yq^2/m_t^2}dy \ ,
\end{equation}
and the quark charges are denoted $q_t,q_b$.
The above vertex  is further connected to the lepton propagator to
produce EDM (see Fig.~1).

\begin{equation}
d_e=\left({3eg^2\over 32\pi^2}\right)
    \left({g^2\over 32\pi^2 M_W}\right)
    \left({m_e \over M_W}\right)
    {\rm Im} (c_t^*c_e)
    \left(q_t F_t + q_b F_b\right) \ .
\end{equation}
Here the form factors $F_{\alpha}$ for $\alpha=t,b$ are given by
\begin{equation}
F_\alpha=\int_0^\infty  \int_0^1  {m_t^2  f_\alpha(2-y)
\; dy \,Q^2 dQ^2   \over
    (M_{H^+}^2+Q^2)(m_t^2+yQ^2)(M_W^2+Q^2)} \ ; \quad f_t=(1-y)\ , \quad f_b=y\ .
\end{equation}

The integrations can be carried
through analytically,
\begin{equation}
T(z)={1- 3 z  \over z^2}{\pi^2 \over 6}
    -\left({1\over z}-{5\over2}\right){\rm ln\ } z
    -{1\over z}
    -\left(2-{1\over z}\right) \left(1-{1\over z}\right)
    {\rm  Sp}(1-z)
    \ ,
\end{equation}
\begin{equation}
B(z)={1\over z}+{2 z -1 \over z^2}{\pi^2 \over 6}
    +\left({3\over2}-{1\over z}\right){\rm ln\ } z
    -\left(2-{1\over z}\right) {1\over z}{\rm  Sp}(1-z) \ ,
\end{equation}
\begin{equation}
F_t={T(z_H)-T(z_W) \over z_H - z_W } \ , \quad
F_b={B(z_H)-B(z_W) \over z_H - z_W }  \ ,
\end{equation}
with
$z_H=M_{H^+}^2/m_t^2$ and $z_W=M_W^2/m_t^2 $,
and the Spence function is defined by
${\rm Sp}(z)=-\int_0^z t^{-1}{\rm\ln}(1-t)dt$
with the normalization ${\rm Sp}(1)=\pi^2/6 $.

\section*{Models}
\quad\vskip -1cm

To illustrate how easily this
mechanism can give rise to a measurable electron EDM,
we consider some specific models.
We first note that the
charged Higgs contribution to $d_e$ vanishes to two loops in the
much-studied simple extension of the Standard Model ---
the two Higgs-doublet Model (2HDM)\cite{Model2}, with the softly broken
discrete symmetry\cite{Glashow&Weinberg} imposed
to enforce natural flavor conservation (NFC).
At one loop level single charged
Higgs exchange between fermions does not
violate CP\cite{Steve90}.  As for the two loop contribution, setting the
scalar doublets $\phi_i=(\phi^+_i,\phi^0_i),i=1,2$ to have
purely real VEVs $\langle\phi^0_i\rangle_0 = v_i$,
in the unitary gauge the charged Higgs propagators
$ \langle \phi^+_1\phi^{+\dag}_2\rangle_0,
  \langle \phi^+_1\phi^{+\dag}_1\rangle_0,
  \langle \phi^+_2\phi^{+\dag}_2\rangle_0 $
are purely real\cite{Steve89,Steve90}.
Ignoring
the CKM matrix, the {\em only}\cite{Steve90} complex coupling  and thus
CP phase in the
lagrangian involving the charged Higgs appears in
${\rm Re}[h(\phi_1^{\dag}\phi_2-v_1 v_2)^2]$;
thus  charged Higgs CP violation in this model
necessarily also involves the pseudoscalar neutral Higgs.

There are, however, several other simple models which can easily contain
sufficient CP violation to produce an electron EDM at an
observable level, through exchange of a charged Higgs boson between fermions.

\noindent
\underline{2HDM III.}

In Model III with two Higgs doublets\cite{Soni,Wolf}, we can choose a basis
so that
$\langle\phi^0\rangle={v\over \sqrt{2}}$ and
$\langle\phi'^0\rangle=0$.
Then $\phi$ mimics the SM Higgs, and $\phi'$ produces new
physics beyond the SM; the physical charged Higgs  $H^+$ is just
$\phi'^+$.
The Yukawa Lagrangian for Model III is
\begin{equation}
-{\cal L}_Y = \eta_{ij}^U \bar Q_{iL} \tilde{\phi} U_{jR} +
              \eta_{ij}^D \bar Q_{iL} \phi D_{jR} +
       \xi_{ij}^U \bar Q_{iL} \tilde{\phi'} U_{jR} +
       \xi_{ij}^D \bar Q_{iL} \phi' D_{jR}
\end{equation}
$$\quad\quad           +\eta_{ij}^E \bar L_{iL} \phi E_{jR} +
       \xi_{ij}^E \bar L_{iL} \phi' E_{jR}   \quad \; + {\rm H.c.}
$$
Here $i,j$ are generation indices. Coupling matrices $\eta$ and
$\xi$ are, in general, non-diagonal.
$Q_{iL}$ and $L_{iL}$ are the
left-handed SU(2) doublets for quarks and leptons.
$U_{jR}$, $D_{jR}$, and $E_R$ are the right-handed SU(2) singlets for
up-type and down-type quarks and charged leptons respectively.
$\langle\phi\rangle$ generates all fermion mass matrices which are
diagonalized by bi-unitary transformations, {\it e.g.}
$M_U={\rm diag}(m_u,m_c,m_t)= {v\over \sqrt{2}}
({\cal L}^U)^{\dagger} \eta^U (\cal R^U)$.
In terms of the mass eigenstates $U$, $D$, $E$, and  $N$(neutrinos), the
relevant
charged Higgs interaction is given by

\begin{equation}
{\cal L}_{H^+} =
    -  H^+ \bar U \biggr[ V_{\rm KM} {\hat \xi}^D
             \hbox{$1\over2$} (1+\gamma^5) -
     {\hat\xi}^{U\dagger} V_{\rm KM}
             \hbox{$1\over2$}(1-\gamma^5) \biggr] D
    -  H^+ \bar N
       {\hat\xi}^E
             \hbox{$1\over2$}
 (1+\gamma^5)  E  + {\rm H.c.}\;\;,
\label{rule}
\end{equation}
with
the CKM matrix
$V_{\rm CKM}=({\cal L}_U)^{\dagger}({\cal L}_D)$, and
$\hat\xi^{P} = ({\cal L}^{P})^{\dagger} \xi^{P} ({\cal R}^{P})$ (for
$P=U,D,E$).

Tree-level flavor changing neutral currents (FCNC)
 are implied by non-zero off-diagonal elements of the
 matrices $\hat\xi^{U,D,E}$.
We adopt the simple ansatz\cite{Soni}
$
\hat\xi^{U,D,E}_{ij} = \lambda_{ij} {g\sqrt{m_i
m_j}}/({\sqrt{2}M_W})
\ .
$
The mass hierarchy ensures that FCNC within the first two
generations are naturally suppressed by  small quark masses, while
a larger freedom is allowed for  FCNC involving the third
generations.  Here $\lambda_{ij}$ can be
$O(1)$ and complex.
CP is already not a symmetry even if we restrict our attention to the
flavor conserving diagonal entries of $\lambda_{ii}$.
We consider only the  third generation quark contribution and set $(V_{\rm
CKM})_{tb}=1$.
The parameters $c_t$
and $c_e$ in Eq.(\ref{eq:lag}) are given as,
$
c_e=-\lambda_{ee} \ , \quad c_t=\lambda_{tt}^* \ .
$
While $\lambda_{tt}$ cannot be significantly larger than $O(1)$ without
producing strong coupling to the top quark, clearly $\lambda_{ee}$ can
be much larger than $O(1)$ so that ${\rm Im} (c_t^*c_e) \gg 1$ is quite
allowed.

\noindent
\underline{3HDM.}

Charged Higgs  sector CP violation can occur in the
three Higgs doublet model\cite{chmod,ch-nedm}.
The first two doublets $\phi_1$ and $\phi_2$ are responsible for the
masses of the $b$--like quarks and the $t$--like quarks respectively.
The charged leptons $e,\mu$ and $\tau$ only couple to $\phi_1$.
The last doublet $\phi_3$ does not couple to the known fermions.
This assignment naturally preserves NFC.
The mass eigenstates $H_1^+$ and $H_2^+$ together with the
unphysical charged Goldstone boson $H_3^+$ are linear combinations of
$\phi_i^+$:
$
\phi_i^+=\sum_{j=1}^{3} U_{ij} H_j^+  \quad(i=1,2,3)
$.
As with the CKM matrix for three quark generations in the SM,
the mixing amplitude $U_{ij}$ matrix generally
contains a single non-zero complex phase,
which  gives rise to  CP nonconservation through the
Yukawa couplings,
$ c_t= U_{2i} (v/v_2) , \;
c_e= U_{1i}  (v/v_1)
$.
In the approximation that  the lightest Higgs dominates, the index $i$
refers to the lightest charged Higgs. As with the 2HDM III, ${\rm
Im} (c_t^*c_e)$ can be much larger than one --- which occurs here if
$v_1 \ll v_2$  (the possibility $v_2 \ll v_1$ is constrained by
maintaining perturbative coupling to the top quark).

%
\section*{Discussion}
\quad\vskip -1cm

We have analyzed the heavy fermion contribution
to the electron EDM due to CP
violation in the charged Higgs sector. From the general structure of the
typical models discussed above, we have shown that the relevant CP
violating parameter Im$(c_t^*c_e)$ can be of order one or larger.  In
Fig.~2, we show the dependence of the electron EDM on $M_{H^+}$ for the
case Im$(c_t^*c_e)=1$. The size of $d_e$ is naturally around $10^{-26}$
e$\cdot$cm, around the current limit.
As noted above, in any specific model there may be other
contributions to $d_e$, but in the absence of accidental
strong cancellation, we expect that the results
presented here reasonably estimate a {\em lower limit} on $d_e$
which applies to a very wide class of CP violating models.
In such case, Fig.~2 may
be used, for example, to rule out $m_{H^+} > 200\,{\rm GeV}$ for
Im$(c_t^*c_e)\ge 1$.

The muon EDM can be easily obtained by the replacements $m_e \to m_\mu$
and Im$(c_t^*c_e) \to $ Im$(c_t^*c_\mu)$. In this case, the analogous
case of 
Im$(c_t^*c_\mu) = 1$   would lead to
an observable muon EDM,
assuming the proposed future sensitivity down to 
$d_\mu =10^{-24}$~e$\cdot$cm.

Finally, we can carry over the calculation for $d_e$ to estimate
contributes to $d_n$, using SU(6) relations \cite{quarkapprox}:
$d_n = {1\over 3}(4 d_d - d_u)$,
where we obtain the down and up quark EDMs  with replacements as made
for $d_\mu$, but with an additional factor $\eta_q$ multiplying both
$d_d$ and $d_u$ coming from QCD evolution of the quark mass and the
quark EDM \cite{polchinski}:
$ \eta_q =
     q(m_t,m_b)^{16/23}
    q(m_b,m_c)^{16/25}
    q(m_c,\mu)^{16/27},
$ with $q(m_a,m_b)= \alpha_s(m_a^2)/ \alpha_s(m_b^2)$.
There are  sizable uncertainties coming from the quark masses and the
extraction of $d_n$ from $d_d$ and $d_u$, but the resulting neutron EDM
should be $d_n \approx 10^{-27}\,(m_d / m_e) {\rm Im}(c_t^*c_d) \,{\rm
e}\cdot{\rm cm}$ for $m_{H^+} \approx 100$ GeV (ignoring the up quark
contribution). This contribution would reach the observable limit for
${\rm Im}(c_t^*c_d) > 6$. In contrast to the case of $d_e$ or $d_\mu$,
there is a sizable contribution from the charged Higgs boson through the
three-gluon operator \cite{CKLY,NIR}. The relative magnitudes are highly
model-dependent: in 2HDM III, the three-gluon operator may vanish even
while the two-loop contributions presented here are non-zero, if $c_t$
is purely real but $c_e$ remains complex. In the 3HDM,  however, the two
contributions to $d_n$ are either both zero or both  non-zero. In any
case, barring strong cancellations, our result places a limit of
$m_{H^+} > 100 $ GeV for ${\rm Im}(c_t^*c_d) =6$.

D.~B.-C. and W.-Y.~K. are supported by a grant from the Department of
Energy, and
D.~C. by a grant from the National Science Council of R.O.C. We also wish to
thank I.~Phillips, T-C.~Yuan and C.-S.~Li who were
involved in an earlier stage of this work.

\section*{Figures}
\begin{itemize}

\item[Fig. 1.]  A typical two-loop Feynman diagram for the electron EDM
due to  charged Higgs sector CP violation. The other diagrams for the
one-loop subgraph $H^-\rightarrow W^-\gamma$ may be found
in Ref.\cite{Raychaudhuri}.

\item[Fig. 2.]
Model-independent contributions to $d_e$ versus $M_{H^+}$ for
Im$(c_t^*c_e)= 1/2, \,1$. The horizontal line denotes the current
experimental limit.
\end{itemize}


\def\baselinestretch{1.2}

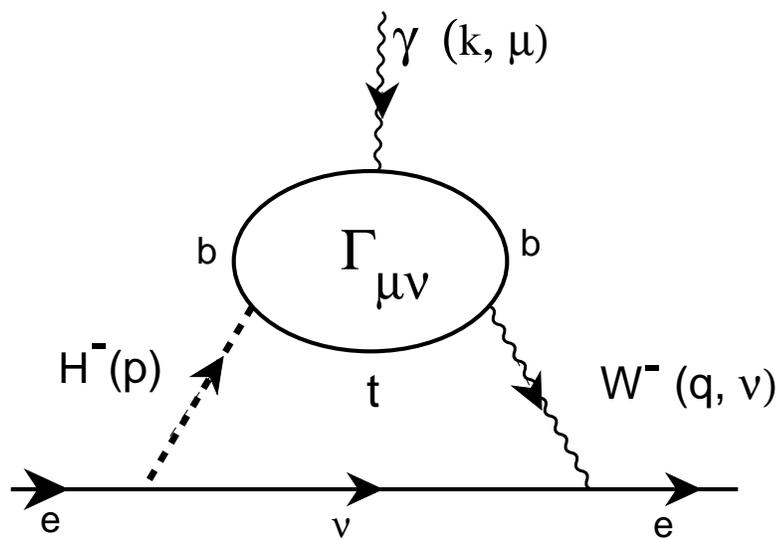
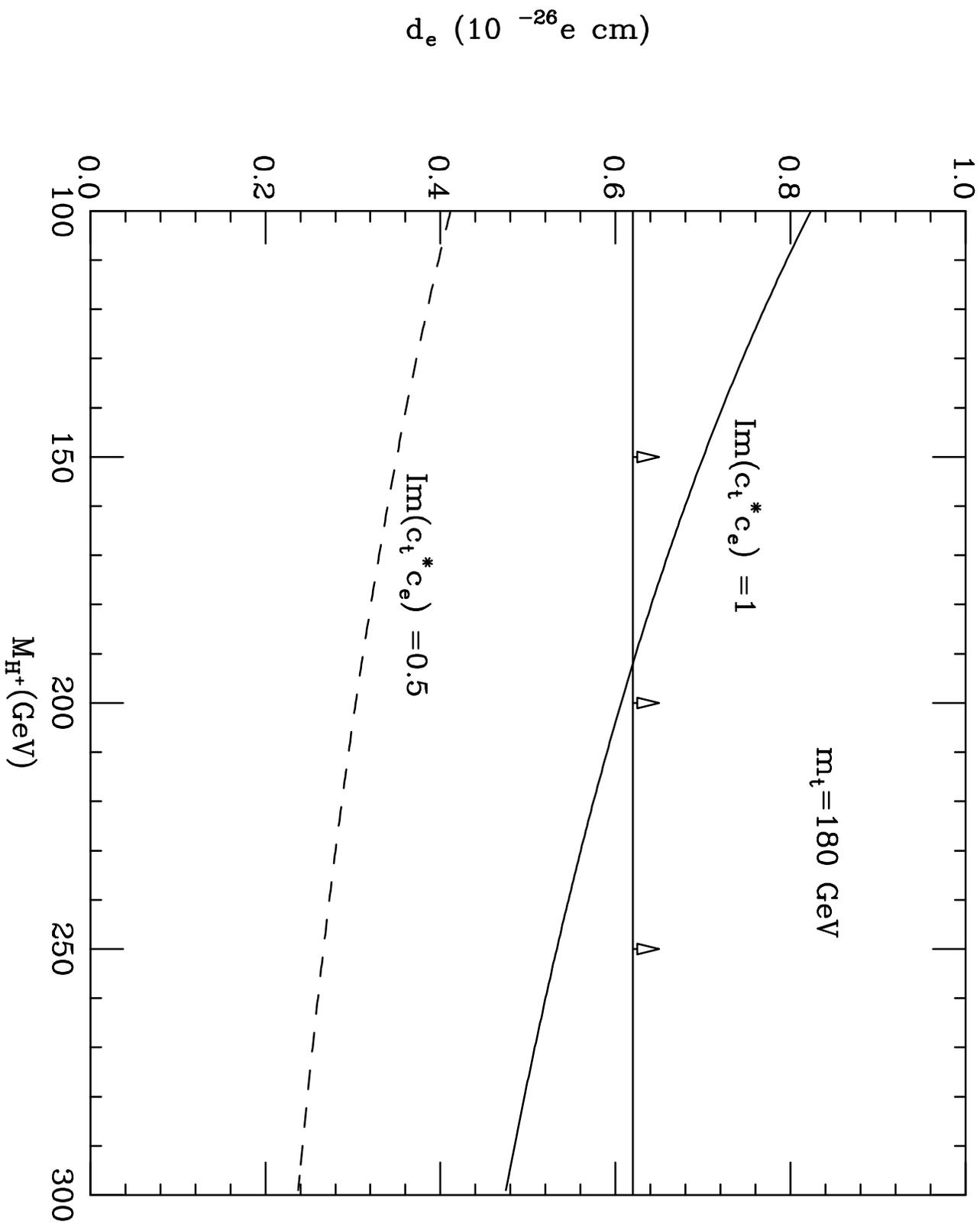
\end{document}